# SURFACE STATES AND RESONANCES IN FIELD EMISSION FROM LOW INDEX FACETS OF CLEAN TUNGSTEN

Z.A. Ibrahim and M.J.G. Lee

Department of Physics, University of Toronto, 60 St George St., Toronto, ON, M5S 1A7, Canada

## Abstract

The energies, widths, and shapes of features observed in the total energy distributions in field emission from W(100) and W(111) are compared with the results of a full-potential LAPW calculation of the surface density of states based on a supercell model of the crystal structure at the metal-vacuum interface. The Swanson hump on W(100) is attributed to two bands of surface states and surface resonances of $d_{z^2}$ symmetry that are highly localised at $\bar{\Gamma}$, and a second peak observed at lower energy is attributed to a band of surface resonances, also of $d_{z^2}$ symmetry, centred at 0.11 Å$^{-1}$ from $\bar{\Gamma}$ along $\bar{\Gamma}\bar{X}$. The energy scale of the calculated total energy distribution is compressed by about 20% relative to the experimental data. The present calculation yields strong evidence that the broad asymmetric peak observed on W(111) is due to emission from a band of surface resonances. Further calculations for W(111) are proposed both to test the accuracy of the band model and to take into account the velocity factor that enters in a calculation of the emission current.

# 1. Introduction

The electronic structures of the low index surfaces of tungsten have been the subject of many investigations. As a refractory metal, tungsten is an excellent material for experimental studies on well-characterized surfaces because clean annealed facets can readily be prepared by flashing to white heat. Stimulated by the availability of high quality experimental data showing a rich variety of features, W(100) has been widely adopted as a prototype for theoretical studies of the electronic structures of metal-vacuum interfaces. Moreover, the electronic structure of the tungsten-vacuum interface (especially with adsorbed overlayers) is of great practical importance, as thermionic emission from a heated tungsten filament is widely used in electronic devices.

The total energy distributions of the field emission currents from several low-index facets of tungsten have been measured at 78K [1, 2, 3]. The W(100) data show a prominent peak (the *Swanson hump* [4]) at about –0.35eV (0.35 eV below the Fermi energy $E_f$), and weaker peaks at about –0.75eV, –1.2eV, and –1.5eV, while the W(111) data show peaks at about –0.75eV and –1.4eV. The origins of the Swanson hump and the second peak on W(100) have been the subject of much discussion [5]. Plummer and Gadzuk attributed the Swanson hump to surface state emission [1]. Linearized augmented-plane-wave calculations by Posternak *et al.* [6] confirmed the presence of a surface state at $\bar{\Gamma}$ at about –0.3eV with 93% of its charge density localized in the surface layer. They attributed the second peak to a pair of surface resonances at about –0.8eV, that exist along $\bar{\Gamma}\bar{X}$ and $\bar{\Gamma}\bar{M}$, but vanish at $\bar{\Gamma}$. The energies and charge density distributions of surface states and resonances on tungsten have received much attention [7, 8, 9, 10].

Penn and Plummer [11] have shown how the shapes of features observed in the experimental total energy distribution in field emission can be analyzed to yield information about the density of states in the surface layer. Modinos [12] has reported a calculation of the enhancement factor in field emission from W(100) based on a non-self-consistent potential and neglecting the spin-orbit interaction. He concluded that a strong peak that appears at –0.28eV in a calculation based on a rectangular surface barrier, is shifted to –0.65eV when an image potential is used to model the self-consistent potential outside the metal. Modinos attributed this strong peak to the Swanson hump, but later [13] argued that it corresponds to the second peak in the experimental enhancement factor, and that a feature corresponding to the Swanson hump appears only when the potential is calculated self-consistently. To the knowledge of the present authors, no work has yet been done to account for the features observed in the experimental total energy distribution in field emission from W(111).

The goal of the present work is to account for the energies, widths and shapes of the principal features observed in field emission from W(100) and W(111). The extreme sensitivity of the results of Modinos to the surface potential emphasizes the need for calculations based on a potential that is self-consistent both inside and outside the metal. Since the bulk energy bands of tungsten are strongly perturbed by the spin-orbit interaction, accurate calculations of bulk-like features in the total energy distribution of the emission current require that the spin-orbit interaction be taken into account. The present calculations are based on potentials that are calculated self-consistently both inside and outside the metal, and the spin-orbit interaction is included in the hamiltonian.

LEED studies on clean W(100) have shown that at low temperature a (c2x2) surface structure is formed, while at high temperature (above about 150K) the surface structure is (1x1) [14, 15]. The surface Brillouin zone (SBZ) that corresponds to the low temperature structure is smaller in linear dimension by a factor of √2, and rotated by 45°, relative to that of the high temperature structure. As a consequence of the surface rearrangement, features at $\overline{M}$ in the SBZ of the high temperature structure are folded to $\overline{\Gamma}$ at low temperature, and so will contribute to the structure in the total energy distribution of the field emission current. Surface rearrangement is a possible complication in the interpretation of field emission data taken at low temperature.

Most of the previous theoretical studies of tungsten-vacuum interfaces have made use of thin film models to represent the transition from the surface to the bulk [6, 7, 8, 9]. In the present study, the full-potential linearized augmented plane wave (FP-LAPW) method is used to calculate the surface electronic structures of W(100) and W(111), based on supercell models of the crystal structures of the metal-vacuum interface. The present calculations are carried out semirelativistically and the spin-orbit interaction is included. They predict structures in the surface density of states on W(100) that are in good agreement with the experimental observations. The results are also generally consistent with thin-film calculations, demonstrating the accuracy of the supercell method and supporting its use to treat more complex surfaces, such as W(111), for which no thin-film calculations have been carried out.

The remainder of this paper is organized as follows. In Section 2 the experimental procedure and the computational method are outlined. In Section 3, the experimental data and the results of the calculations are presented, first for W(100) and then for W(111). Finally in Section 4, the conclusions of the present work are summarized.

## 2. Experimental and Computational Techniques

### A. Apparatus and experimental procedure

The measurements were carried out at room temperature in a stainless steel vacuum system at a pressure lower than $10^{-10}$ Torr. The field-emission spectrometer has been described in detail elsewhere [16]. A schematic diagram is shown in Fig. 1. A tungsten field-emitter (the tip) is mounted on a sample holder about 5 cm from a fluorescent conducting screen with a probe hole at its center. A potential difference of several kilovolts is applied between the field emitter and the screen. Electrons that are field emitted from the tip are accelerated towards the screen where they produce a magnified image of work function variations across the surface of the tip. Emission from a particular facet is selected by adjusting the electrostatic potentials on the deflector plates to center the image of the desired facet over the probe hole. Electrons that pass through the probe hole are decelerated by the electrostatic lenses, and the emerging parallel beam enters a double-pass 127° cylindrical energy analyzer. Electrons whose total energies lie within a narrow range about the pass energy emerge through the exit slit and are counted by a Spiraltron electron multiplier.

The total energy distribution (TED) of electrons field emitted from the chosen facet was measured by stepping the bias voltage through 100 channels, each 25 mV wide, spanning the appropriate energy range, the tip-to-screen potential difference being held constant. At each step the bias voltage was allowed to stabilize for 1ms, then the field-emitted electrons were counted for 20ms. The number of cycles (typically 20) was chosen to allow for adequate statistics while avoiding significant contamination of the tip.

## B. Data analysis

Assuming that the potential barrier at the emitting surface varies only in the normal direction, the transmission probability $D(W)$ is a function of the normal energy $W$. The total energy distribution (TED) in field emission is calculated by integrating the charge flux $N(E,W)dE\,dW$ associated with electrons of total energy $E$ and normal energy $W$, multiplied by the transmission probability $D(W)$, over all possible values of $W$:

$$j(E) = \int_0^E dW\, N(E,W)D(W), \qquad (1)$$

where the zero of energy is taken to be at the bottom of the conduction band. $N(E,W)$ depends on the Fermi-distribution function $f(E)$, the group velocity of the electrons, and the density of electronic states [17].

In the free electron approximation, Eq. (1) reduces to:

$$j_o(E) = (4\pi me/h^3)f(E) \int_0^E dW\, D(W), \qquad (2)$$

where $m$ and $e$ are the mass and the magnitude of the charge of the electron respectively. Taking into account the image potential, and calculating $D(W)$ in the WKB approximation, the free-electron TED takes the following form.

$$j_o(E) = A\, exp[(E-E_f)/d]\, [1+exp[(E-E_f)/(kT)]]^{-1}, \qquad (3)$$

where $E_f$ is the Fermi energy, $k$ is Boltzmann's constant, $T$ is the absolute temperature, $d$ is a known function of the electric field strength and the work function at the emitting surface, and $A$ is an energy-independent pre-factor [18]. According to Eq. (3), a semilogarithmic plot of $j_o(E)$ against $E$ will be triangular in shape with a peak at the Fermi energy. The peak will be slightly rounded at finite temperatures.

Features in the surface density of states at the metal-vacuum interface show up as deviations from the free-electron model. In order to remove the irrelevant effects of the tunneling barrier and thermal excitation, it is convenient to divide the experimentally-observed TED by the free-electron TED to yield the enhancement factor $R(E)$:

$$R(E) = j(E) / j_o(E). \qquad (4)$$

$R(E)$ is a measure of the energy dependence of the electronic density of states in the surface layer [11,19].

In the present work, the free-electron TED $j_o(E)$ calculated from Eq. (3) was convolved with a Gaussian distribution of width 60 meV to take into account the instrumental resolution function $\Delta(E)$. The Fermi energy and the strength of the electric field at the emitting surface were adjusted to fit $j_o(E)\otimes\Delta(E)$ to the background of the experimentally-observed TED. The enhancement factor $R(E)$ was extracted from the best fit to the data according to Eq. (4). The work functions assumed in the calculation of the free-electron distribution are 4.64eV for W(100) and 4.45eV for W(111) [20].

## C. Computational details

The electronic structures of the W(100)-vacuum and W(111)-vacuum interfaces have been calculated semirelativistically within the framework of density functional theory. The spin-orbit interaction was included in all of our calculations except where otherwise noted. Exchange and correlation were taken into account in the generalized gradient approximation (GGA) [21]. The interface was represented by a supercell containing 13 layers of tungsten atoms and an atom-free region of equal volume to represent the vacuum. The electronic structure was calculated self-consistently using the full-potential linearized augmented plane wave (FP-LAPW) method. A comparison between the densities of states in the central layers and the bulk density of states indicates that on W(100) a supercell of 13 layers adequately represents the transition from the surface to the bulk.

The charge distribution associated with each electron state was decomposed layer by layer. Those states whose charge densities are enhanced by at least 30% in the two outermost layers of the supercell were classified as surface resonances. Those states having at least 75% of their total charge in the surface layer and also a surface-to-bulk charge ratio in excess of 100 were classified as surface states. In each layer, the charge density of each electron state was decomposed into components corresponding to the symmetry elements of the point group. Because the surface potential barrier selects strongly for electrons that propagate almost normal to the emitting surface (the z direction), only the *s, $p_z$, $d_{xz}$, $d_{yz}$,* and *$d_{z^2}$* components contribute significantly to field emission. Only those surface states and surface resonances that have the appropriate symmetries to contribute to field emission are shown in the dispersion curves presented in this paper.

Our calculations were carried out using the WIEN97 implementation of the full-potential linearized augmented plane wave (FP-LAPW) method [22]. In the distributed program, each electron state is represented by its *k* vector in an irreducible sector of the Brillouin zone. The program was modified to keep track of its full *k* vector. This makes it possible to calculate the normal energy and hence the tunneling factor associated with each electron state. To facilitate comparison with experimental data, the results of the present calculations are reported as k-resolved layer densities of states (K-LDOS). The k-resolved layer density of states at the surface is denoted K-SDOS.

The K-LDOS is calculated by multiplying each contribution to the layer density of states by the tunneling factor and the Fermi factor, and dividing by the free-electron TED $j_o(E)$ (the Fermi factor cancels out). It is important to note that the calculated K-LDOS, which contains no velocity factor, is a k-weighted density of states, while the experimental enhancement factor is a k-weighted current. Thus a feature in the calculated K-LDOS will be directly comparable to the corresponding feature in the experimental enhancement factor *R(E)* only if it is dominated by contributions from a region of the surface Brillouin zone over which the velocity factor varies little. In particular, a band of surface states that is highly localized in the surface Brillouin zone is expected to yield similar features in the two distributions, while a broad band of surface resonances is expected to yield a more extended feature in *R(E)*, where a decreasing surface density of states tends to be compensated by an increasing velocity factor [23].

## 3. Results and Discussion

**A. W(100)**

**(i) Experimental studies of surface states and surface resonances.** In the semilogarithmic plot in Fig. 2, the total energy distribution *j(E)* in field emission from W(100) is compared with the free-electron distribution $j_o(E)$. *j(E)* deviates significantly from $j_o(E)$ in the energy range between $E_f$ and 1.0eV below $E_f$. The low and the high-energy tails observed outside this energy range are due to the scattering of electrons at the walls of the energy analyser [24].

Fig. 3c shows the enhancement factor *R(E)* extracted from the experimental TED. The principal feature is a symmetric peak, the Swanson hump, centered at –0.36eV (0.36eV below $E_f$) and having a full width at half maximum height (FWHM) of 0.22eV. At about -0.66eV a second poorly-resolved feature is superimposed on the tail of the Swanson hump. The second peak has previously been observed in several low temperature studies on clean W(100), but not to our knowledge at room temperature. The present results are evidence that the second peak is not induced by the surface rearrangement that may occur at low temperature on the (100) facet of a tungsten field emitter [14, 15].

The energies of features observed in the various experimental measurements of the enhancement factor for field emission from W(100) are compared in Table I. In the present work, the presence of the second peak was inferred by subtracting a Lorentzian fit to the Swanson peak from the experimental enhancement factor. As the second peak is not observed directly, the earlier determinations of the energy of this peak are considered to be more reliable. The data of Ref. 2 show a weak peak that extends from –1.1eV to –1.3eV. Over this energy range the data of Ref. 1 indicate that the enhancement factor is constant, so this third peak appears to represent a feature of the TED that was not resolved in the data of Ref. 1. Below –1.3eV a low-energy tail dominates the data of Ref. 2, while the data of Ref. 1 show a fourth peak at –1.5eV.

Table I. Energies of the principal features of the TED for field emission from W(100).

| Feature | Energy below $E_f$ [eV] | | | | |
| --- | --- | --- | --- | --- | --- |
| | Experiment | | | Calculation | |
| | Ref. [1] | Ref. [2] | Present | Ref. [13] | Present |
| Swanson peak | 0.37 | 0.33 | 0.36 | 0.27 | 0.31 |
| Second peak | 0.78 | 0.73 | 0.66 | 0.85 | 0.63 |
| Third peak | - | 1.20 | - | - | 0.95 |
| Fourth peak | 1.50 | - | - | 1.50 | 1.22 |

**(ii) Interpretation of the experimental data.** Figure 3b shows the calculated k-resolved surface density of states (K-SDOS) from W(100). The calculated peak *A* is centered at about –0.31eV and has a FWHM of 0.20eV. The width and shape of the calculated peak are in good agreement with those of the Swanson hump. The agreement between the shapes suggests that the Swanson hump is due to emission from electron states in a highly localized region of the surface Brillouin zone over which the velocity factor varies little.

The dispersion curves for surface states and surface resonances along $\overline{\Gamma}\,\overline{X}$ are shown in Fig. 3a. The peak in the K-SDOS (labeled *A*) arises from two bands, both of $d_{z^2}$ symmetry, that extend over a narrow region of the surface Brillouin zone close to $\overline{\Gamma}$. Very close to $\overline{\Gamma}$ the electron states in the two bands are surface states (marked by arrows) with energies of –0.28eV and –0.32eV. The splitting between these two surface states vanishes in the absence of spin-orbit interaction. The Swanson hump is attributed to emission from these two bands of surface states and surface resonances. Other authors have also attributed the Swanson hump to surface state emission [1, 6, 7, 8, 9, 25]. In Fig. 4 the calculated K-SDOS is compared with the K-LDOS in the second layer and at the center of the supercell. Peak *A* is weak in the second layer, and is not resolved as a distinct peak in the bulk. This indicates that the surface resonances that contribute to peak A are strongly enhanced in the surface layer.

It is of interest to examine how the states that contribute to the Swanson hump are distributed throughout the surface Brillouin zone. Contours of constant K-SDOS in the vicinity of $\overline{\Gamma}$ at –0.28eV and –0.32eV are plotted in Fig. 5a and Fig. 5b respectively. The plots show that most of the contribution comes from a region of the surface Brillouin zone centered on $\overline{\Gamma}$ and of radius $0.1\,\overline{\Gamma}\,\overline{X}$ (~0.1 Å$^{-1}$).

The K-SDOS in Fig. 3b shows a second peak, labeled B, superimposed on the low energy tail of peak A and centered at about –0.63eV. The dispersion curve shows that this feature arises from a band of surface resonances of $d_{z^2}$ symmetry. This peak is close in energy to the poorly resolved feature that was observed by experiment. The K-LDOS plot in Fig. 4 shows that peak B extends to the bulk. As can be seen from the contours of constant K-SDOS shown in Fig. 5c, the peak in the K-SDOS at –0.63eV is due to a region of surface resonances centered at $0.11\overline{\Gamma}\overline{X}$, and states at $\overline{\Gamma}$ make no contribution.

The K-SDOS shows a weak peak, labeled E, at about –0.95eV. The layer densities of states in Fig. (4) show that the charge density associated with this peak extends to the bulk. This third peak arises from the contributions of two bulk bands of $d_{z^2}$ and $d_{xz,yz}$ symmetry in a highly-localized region of the surface Brillouin zone centered at about $0.11\overline{\Gamma}\overline{X}$. A fourth peak, labeled F, is centered at about –1.2eV. This peak is due to an intermediate band, predominantly of $d_{xz,yz}$ symmetry, with a peak charge density in the second layer.

In Table I, the energies of the principal features of the calculated TEDs for field emission from W(100) are compared with the experimental data. While the present calculation accounts for all of the features of the experimental data, the energy scale of the calculated TED is contracted by about 20% relative to the experimental data. The features observed in the enhancement factor calculated by Modinos [13] are also in good overall agreement with the experimental data. Discrepancies in the energies of features in the two calculations are to be expected, both because the energies are very sensitive to the potential outside the metal [12] where Modinos used an image potential to approximate the self-consistent potential, and because the spin-orbit interaction was not included in his calculation.

**(iii) Role of spin-orbit interaction.** The K-SDOS in Fig. 3b shows a peak *C* centered about 0.03eV above $E_f$ and a shoulder *D* about 0.50eV below $E_f$, both of which are induced by the spin-orbit interaction. Both features are surface resonances of $d_{xz,yz}$ symmetry, and while they are strongly enhanced in the surface layer, they also have appreciable strength in the second layer and in the bulk, as shown in Fig. 4.

The spin-orbit interaction induces an energy gap in the bulk states along the (100) direction of tungsten. Fig. 6 shows the dispersion curves along $\overline{\Gamma}\,\overline{H}$ and the corresponding K-SDOS resulting from two different calculations of the bulk-band structure, one neglecting the spin-orbit interaction (Fig. 6a) and the other including it (Fig. 6b). The strong peak at –0.31eV and the weaker peaks at –0.63eV, –0.95eV, and –1.22eV all appear irrespective of whether the spin-orbit interaction is included in the hamiltonian, confirming that none of these peaks is induced by the spin-orbit interaction [13, 25]. According to the present calculations, none of these peaks shifts in energy by more than about 0.1eV when the spin-orbit interaction is turned on.

Peak *C* and shoulder *D* coincide in energy with the edges of the spin-orbit gap, and neither appears unless the spin-orbit interaction is included in the calculation. The fact that no lens orbit is seen in experimental Fermi surface data for tungsten is evidence that the upper edge of the spin-orbit gap (peak *C*) lies above the Fermi level. This may explain why no peak corresponding to *C* appears below the Fermi level in the experimental TED. Both peak *B* and shoulder *D* are expected to contribute to the second peak observed experimentally on W(100).

## B. W(111)

**(i) Surface resonances.** In the semilogarithmic plot in Fig. 7, the total energy distribution $j(E)$ from the W(111) facet is compared with the free-electron distribution $j_o(E)$. In the energy range from 0.9eV below $E_f$ to $E_f$, $j(E)$ deviates significantly from $j_o(E)$. The low and the high-energy tails outside this energy range are due to electron scattering at the walls of the energy analyzer [24].

The experimental enhancement factor in Fig. 8c shows a highly asymmetric peak that extends from about –0.9eV, reaches a maximum at about –0.7eV, decays slowly with increasing energy and disappears abruptly at about –0.18eV. The calculated K-SDOS in Fig. 8b shows a broad asymmetric peak (labeled *A*) in the same energy range, that is due to a band of surface resonances of $d_{z^2}$ symmetry as shown by the dispersion curves in Fig. 8a. The experimental peak disappears abruptly close to where the calculated band loses surface resonance character. This observation supports the interpretation of the asymmetric peak on W(111) as being due to emission from a band of surface resonances. The peak in the calculated surface density of states is much narrower than the experimental peak. As discussed above, this is consistent with emission from an extended band of surface resonances over which there is significant variation of the velocity factor.

The experimental TED for field emission from W(111) shows no evidence for a feature corresponding to peak *B* in the K-SDOS plot, which is due to a band of surface resonances whose low energy limit is just below the Fermi level. A possible explanation is that the velocity factor approaches zero at the lower energy limit of a band of surface resonances. The low energy limit of the calculated peak *A* is about 0.3 eV above that of the experimental peak. This suggests that, as on (100), the energy scale of the calculated TED is significantly contracted relative to the experimental data.

# 4. Conclusions

The present results for the energies, widths, and shapes of the principal features in the calculated TED for field emission from W(100) are in good overall agreement with the experimental data and with the results of other authors, confirming the accuracy of the present FP-LAPW supercell calculation. The width of the calculated peak in the K-SDOS is in good agreement with the width of the Swanson hump.

The strongly asymmetric peak observed in field emission from W(111) is attributed to emission from a band of surface resonances. However, the calculated peak is significantly narrower than the experimental peak. This suggests that, in a calculation of the emission current on W(111), the variation of the velocity factor over the surface resonance band must be taken into account. We are currently carrying out a calculation of the emission current from a parabolic band of surface resonances in an attempt to account for the width of the peak observed experimentally on W(111). The results of this calculation will be reported elsewhere.

The energy scales of the calculated TEDs for emission from W(100) and from W(111) are significantly contracted relative to the experimental data. This suggests that, over the experimentally accessible energy range, the present method of calculation underestimates the widths of the $d$-like valence bands of tungsten. The present method of calculation also predicts a lens orbit on the Fermi surface of tungsten that is not observed experimentally, indicating that it underestimates the spin-orbit gap along ΓH. These discrepancies, which might well be related, will be the subject of further study.

## Acknowledgements

One of us (Z.A.I.) wishes to acknowledge partial financial support from the University of Toronto. This work was supported in part by a grant from the Natural Sciences and Engineering Research Council of Canada. The authors are indebted to P. Blaha for helpful discussions.

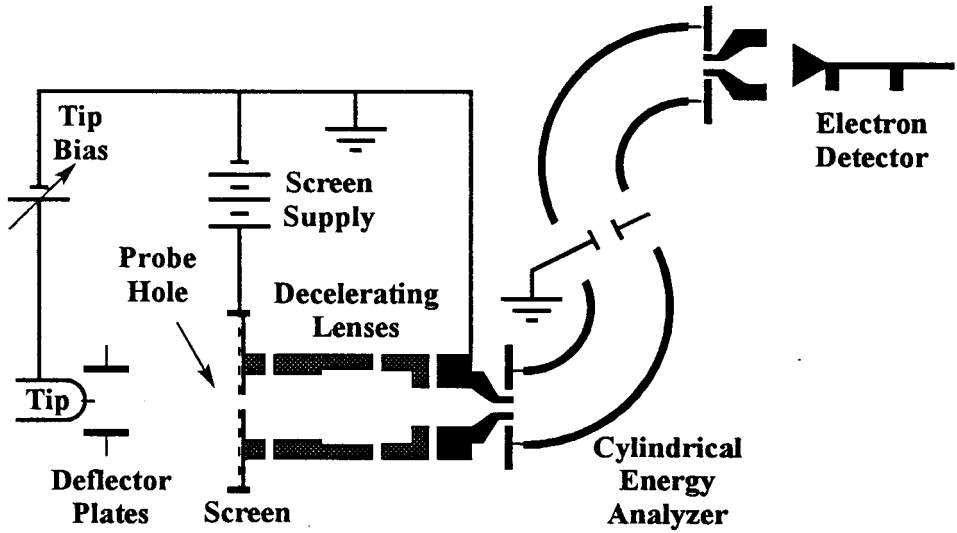

Schematic diagram of the field-emission spectrometer.

Figure 1

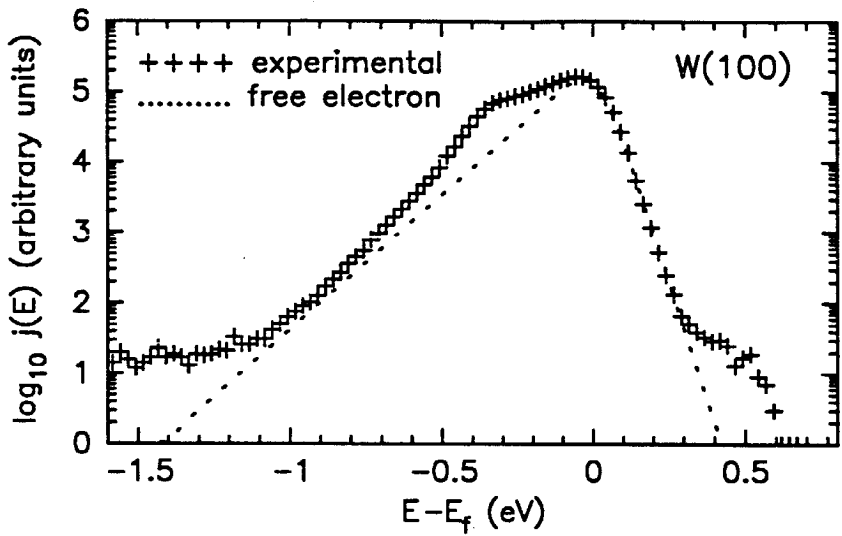

Semilogarithmic plot of the total energy distribution from W(100) compared with the free-electron distribution, showing the extra contribution from the Swanson hump. The field strength deduced from the fit is 0.26V/Å.

Figure 2

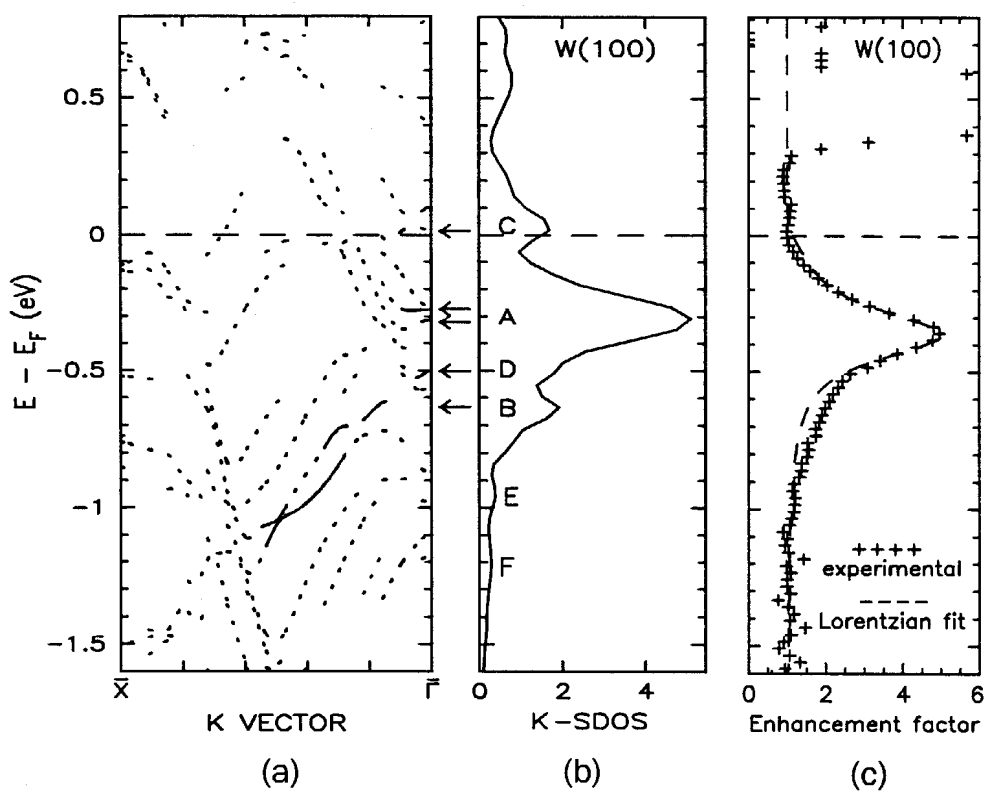

Comparison between the experimental enhancement factor R(E) and the calculated K-SDOS for W(100). (a) Dispersion relation along $\overline{X}\,\overline{\Gamma}$ showing surface states (solid lines) and surface resonances (dotted lines). Arrows mark the surface states and surface resonances that correspond to the Swanson hump and to the other peaks in the calculated K-SDOS. (b) Calculated K-SDOS. (c) Experimental enhancement factor R(E) together with a fitted Lorentzian.

Figure 3

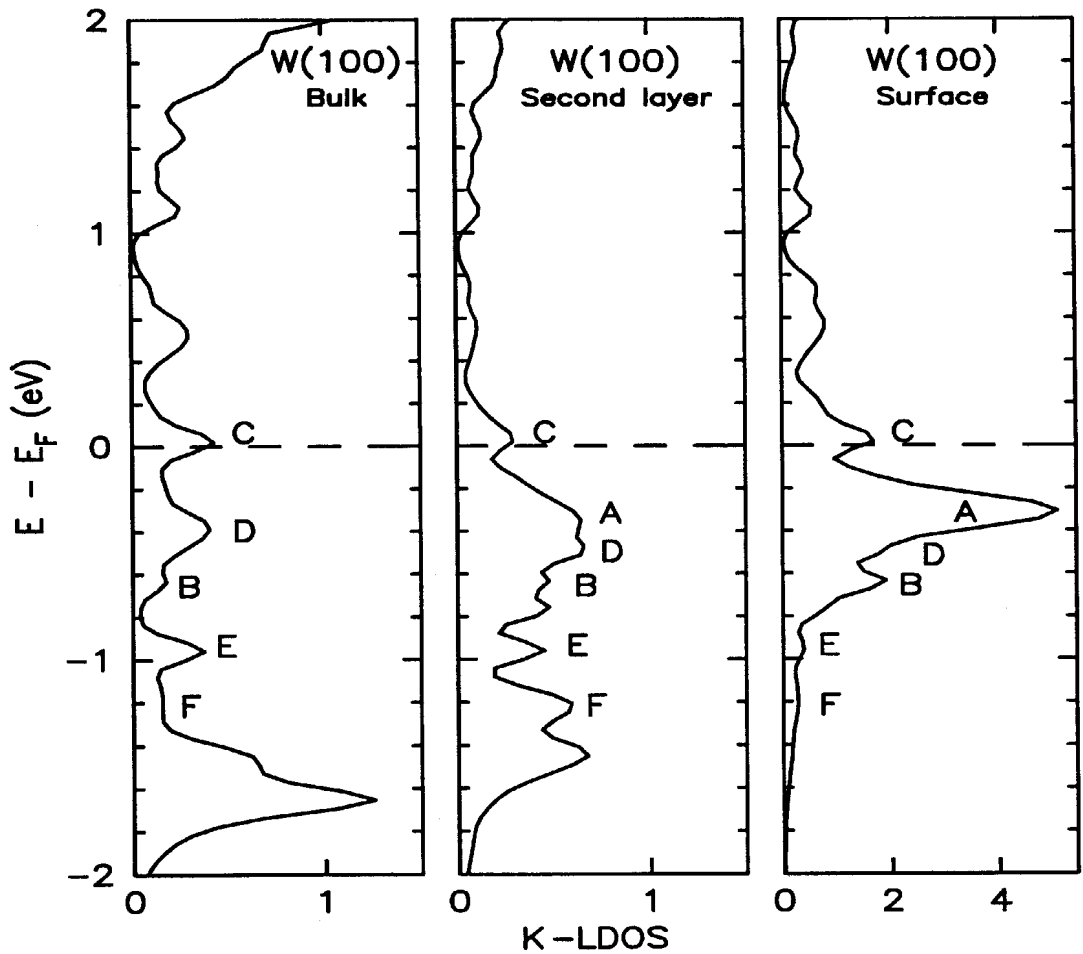

K-LDOS in (a) the bulk, (b) the second layer and (c) the surface of W(100).

Figure 4

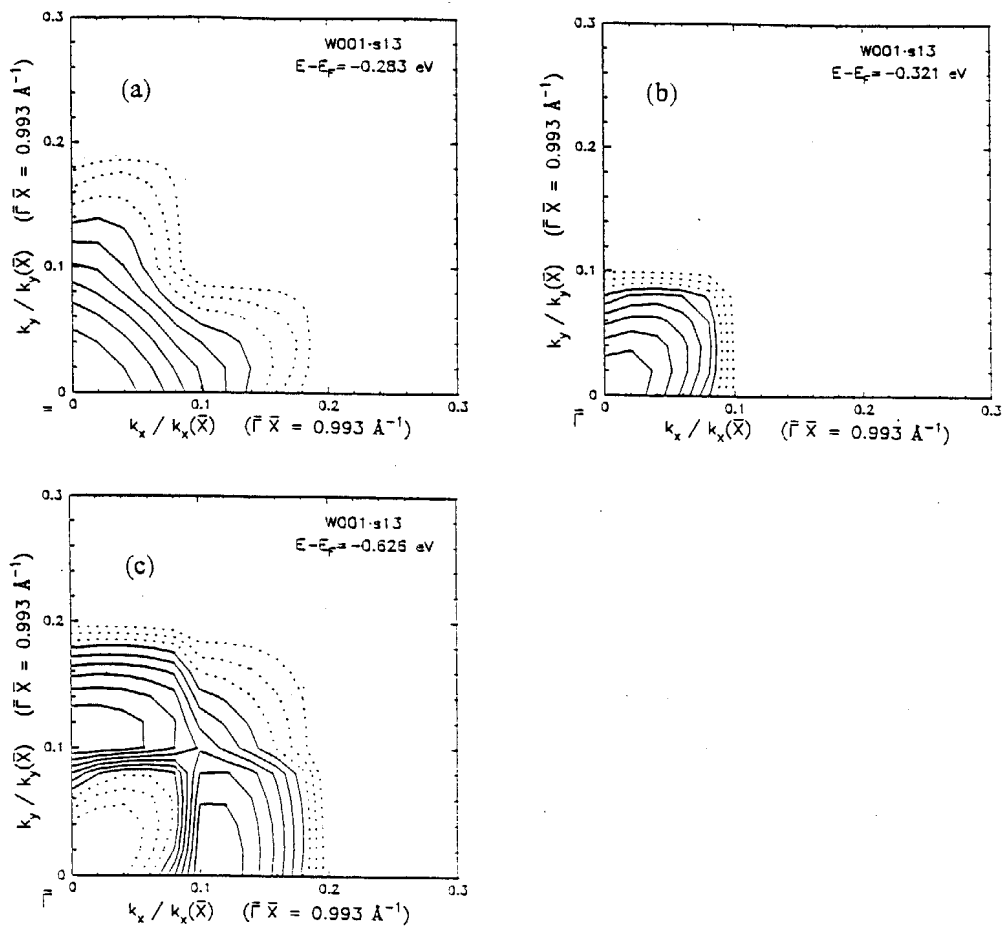

Contours of constant K-SDOS close to $\overline{\Gamma}$. The contours in (a) and (b) at 0.28eV below $E_f$ and 0.32eV below $E_f$ respectively show that the main contributions to the Swanson hump come from a small $\overline{\Gamma}$-centred region of surface states and surface resonances. (c) Peak B at 0.63eV below $E_f$ is due to emission from a small region of surface resonances centred around $0.11\overline{\Gamma}\,\overline{X}$.

Figure 5

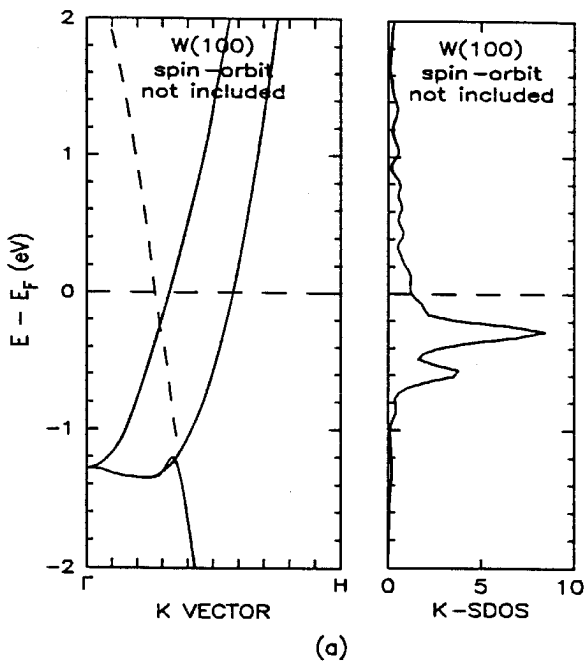

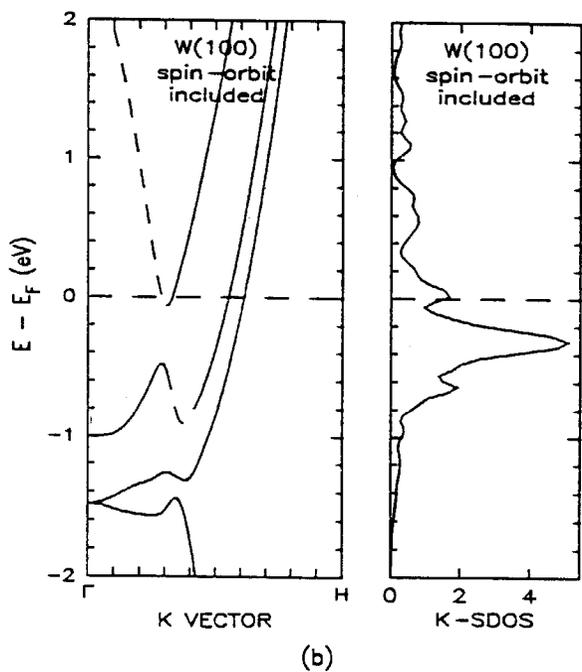

The effect of spin-orbit interaction on the dispersion of the bulk states and on the peaks in the K-SDOS. The spin-orbit interaction is neglected in (a) and included in (b).

Figure 6

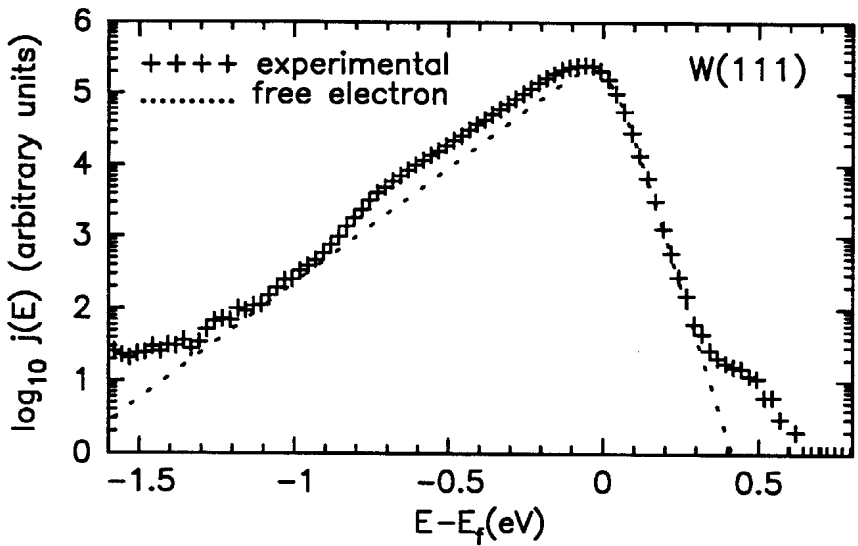

Semilogarithmic plot of the TED of field emission from W(111) compared with the free-electron TED, showing an extra contribution in the form of an asymmetric peak. The field strength deduced from the fit is 0.20V/Å.

Figure 7

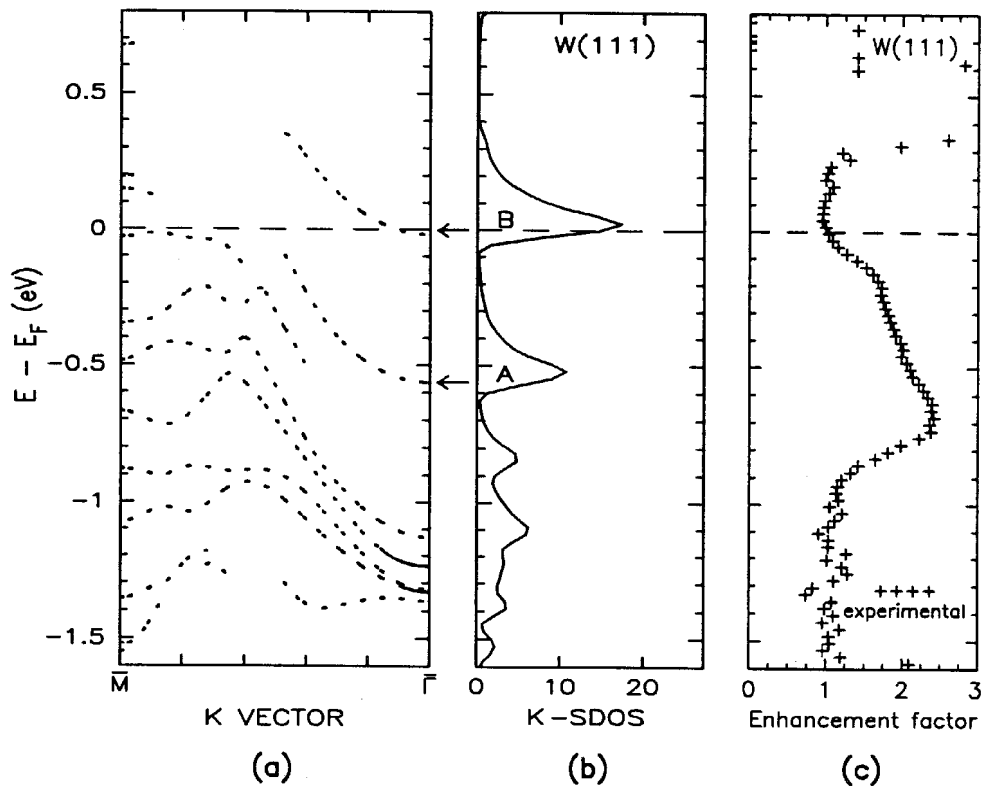

Comparison between the experimental enhancement factor R(E) and the calculated K-SDOS for W(111). (a) The dispersion curves along $\overline{M}\,\overline{\Gamma}$ showing surface states (solid line) and surface resonances (dotted line). The arrows mark the low energy limits of the bands of surface resonances. (b) The calculated K-SDOS. (c) The experimental enhancement factor R(E).

Figure 8